\documentclass[journal]{IEEEtran}
\IEEEoverridecommandlockouts

\usepackage[utf8]{inputenc}
\usepackage[T1]{fontenc}

\usepackage{cite}
\usepackage{blindtext, graphicx}
\usepackage[cmex10]{amsmath}
\usepackage{amsfonts,amsthm,amssymb}
\usepackage[font=scriptsize,caption=false,labelsep=space]{subfig}
\usepackage{graphicx,float,wrapfig}
\usepackage{graphicx}
\usepackage{graphics,color}
\usepackage{mathtools}
\usepackage{nicefrac}
\usepackage{bm}
\usepackage{txfonts}

\def\IEEEbibitemsep{0pt plus .5pt}
\usepackage{balance}

\usepackage[noend]{algpseudocode}
\usepackage{algorithm} 
\usepackage{algorithmicx}   
\usepackage{color}

\usepackage{wrapfig} 
\usepackage{pgfplots}

\usepackage{upquote}
\usepackage[none]{hyphenat}
\pgfplotsset{compat=1.8}

\newtheorem{theorem}{Theorem}

\begin{document}

\setlength{\abovedisplayshortskip}{2pt}
\setlength{\belowdisplayshortskip}{2pt}

\title{TenDSuR: Tensor-Based 4D Sub-Nyquist Radar}


\author{Siqi~Na,
		Kumar Vijay~Mishra,
        Yimin Liu, \IEEEmembership{Member, IEEE,}
        Yonina C.~Eldar, \IEEEmembership{Fellow, IEEE,}
        and Xiqin Wang

\thanks{S.N., Y.L. and X.W. are with the Department of Electronic Engineering, Tsinghua University,
Beijing, China, e-mail: nasiqi123456@163.com, \{yiminliu, wangxq\_ee\}@tsinghua.edu.cn.}
\thanks{K.V.M. and Y.C.E. are with the Andrew and Erna Viterbi Faculty of Electrical Engineering, Technion - Israel Institute of Technology, Haifa, Israel, e-mail: \{mishra, yonina\}@ee.technion.ac.il.}
\thanks{The work of Y. Liu was supported by the National Natural Science Foundation of China under Grant 61571260. K.V.M. acknowledges partial support via Andrew and Erna Finci Viterbi Postdoctoral Fellowship and Lady Davis Postdoctoral Fellowship.}}

\maketitle

\begin{abstract}
We propose Tensor-based 4D Sub-Nyquist Radar (TenDSuR) that samples in spectral, spatial, Doppler, and temporal domains at sub-Nyquist rates while simultaneously recovering the target's direction, Doppler velocity, and range without loss of native resolutions. 
We formulate the radar signal model wherein the received echo samples are represented by a partial third-order tensor. We then apply compressed sensing in the tensor domain and use our tensor-OMP and tensor completion algorithms for signal recovery. 
Our numerical experiments demonstrate joint estimation of all three target parameters at the same native resolutions as a conventional radar but with reduced measurements. Furthermore, tensor completion methods show enhanced performance in off-grid target recovery with respect to tensor-OMP.
\end{abstract}


\IEEEpeerreviewmaketitle

\section{Introduction}
\label{sec:intro}
\IEEEPARstart{I}{n} recent years, radar systems that use fewer measurements in temporal, spectral, Doppler or spatial signal domains and yet achieve identical or even better performance than conventional systems, have captured significant research interest\cite{griffiths2014radar,mishra2018sub,cohen2018sub}. The savings in metrics such as antenna aperture, bandwidth, sampling rate or dwell time lead to low cost and small size systems that are not very demanding on data throughput. Further, reduction in system resources enables the radar to share them among different applications. For instance, technologies like multi-function integrated radio-frequency (RF) aperture systems \cite{Darpa2016Concerto, mishra2017high,mishra2014compressed}, spectral coexistence \cite{cohen2018spectrum,mishra2017auto}, cognitive radars \cite{mishra2016cognitive, mishra2017performance} and reconfigurable arrays \cite{wang2014reconfig} provide multiple services such as surveillance, tracking, and communication using a single system. 

The literature indicates several different approaches towards realizing reduced-rate radars (see e.g. \cite{mishra2018sub,cohen2018sub} and references therein) mostly based on compressed sensing (CS). 
In this work, we focus on those reduced-rate techniques which model the analog received radar signal utilizing the theory of finite-rate-of-innovation (FRI) \cite{eldar2015sampling}. These systems - referred to as sub-Nyquist radars - perform signal detection and parameter estimation from much fewer measurements than that required by Nyquist sampling and employ the Xampling framework \cite{eldar2015sampling} to obtain low-rate samples of the signal.

In the temporal domain, \cite{baransky2014prototype} proposed a sub-Nyquist radar to recover target delays relying on the FRI model. Similar techniques were later studied for delay channel estimation problems in ultra-wideband \cite{cohen2014channel} and millimeter wave \cite{mishra2017sub} communication systems. The \textit{Doppler focusing} technique was added to the temporal sub-Nyquist radar in \cite{barilan2014focusing} to recover both delays and Dopplers. However, this system reduced samples only in time but not in the Doppler domain. The sub-Nyquist reduced time-on-target radar in \cite{cohen2016reduced} demonstrated dilution of samples in both the time and Doppler domains. For spatial compression, thinned arrays were examined for a multiple-input-multiple-output (MIMO) array radar in \cite{rossi2014spatial} and later for phased arrays in \cite{mishra2017high}. In \cite{Cohen2016SUMMeR}, targets' ranges, velocities, and directions  were recovered in a new radar structure called Sub-Nyquist MIMO Radar (SUMMeR) by thinning a colocated MIMO array and collecting low-rate samples at each receiver element. The SUMMeR system was also implemented in a hardware prototype \cite{mishra2018cognitive,mishra2016cognitive}. 

In this work, motivated by the recent advancements in exploring the high-order data structure \cite {Sidiropoulos2017Tensor} in multidimensional harmonic retrieval \cite{Haardt2008Higher}, dictionary learning \cite{roemer2014tensor}, channel sounding \cite{landmann2012impact}, and bistatic MIMO radar processing \cite{wen2017angle}, we propose a Tensor-based four dimensional (4D) Sub-Nyquist Radar (TenDSuR), which is a Spectral-Spatial-Doppler-Temporal (S-S-D-T) domain compressed 4D radar system that processes the received echoes by applying tensor-based signal processing. In TenDSuR, the spatial compression is achieved by deploying a thinned MIMO antenna as in \cite{mishra2017high, Cohen2016SUMMeR}. For the spectral thinning of transmit waveforms as in \cite{Yu2012CSSF, Huang2014Congnitive}, TenDSuR employs frequency-diversity waveforms which occupy only a small part of the full bandwidth required for the range resolution of SUMMeR \cite{Cohen2016SUMMeR}. The sub-Nyquist receiver recovers the target parameters via Xampling leading to temporal compression.

The transmitters send agile waveforms with the reduced number of pulses thereby requiring fewer measurements in the Doppler domain. So, unlike SUMMeR, the TenDSuR system does not transmit pulses at a uniform Pulse Repetition Interval (PRI). This leads to significant reduction in the total duration which each of the antenna elements are dedicated to a specific RF service. Further, SUMMeR employs Doppler focusing, which is carried out over the set of frequencies that are fixed \textit{a priori}. The resultant Doppler resolution is limited by the focusing, i.e., inversely proportional to the number of pulses $P$ as is also the case with conventional radar. In contrast, the TenDSuR processing algorithm is based on tensor completion (TC) \cite{gandy2011tensor} leading to higher resolution recovery of the targets with off-grid range, Doppler, and Direction of Arrival (DoA). The flexible signal model of TenDSuR is suitable for arbitrary waveforms. The results show that in order to achieve the same detection performance with same resolutions in Doppler and DoA, TenDSuR requires fewer pulses per transmitter than SUMMeR \cite{Cohen2016SUMMeR}. When all pulses and frequency points are used, TenDSuR signal model is equivalent to SUMMeR. But, for off-grid targets, TC-based recovery outperforms SUMMeR even when pulses are not diluted.


Throughout this paper, we use bold lowercase, bold uppercase and calligraphic letters for the vectors, matrices and tensors respectively. The $i$th element of a vector $\mathbf{y}$ is $[\mathbf{y}]_i$; the $(i,j)$th entry of a matrix $\textbf{Y}$ is $[\textbf{Y}]_{i,j}$; the $s$th column of matrix $\mathbf{Y}$ is $[\mathbf{Y}]_{s}$; the sub-matrix of $\mathbf{Y}$ that has columns specified by the index $\Pi$ is $[\mathbf{Y}]_{\Pi}$; and the $(i,j,k)$th entry of a tensor $\mathcal{Y}$ is $[\mathcal{Y}]_{i,j,k}$. The notations $\otimes$, $\circ$, and $\diamond$ are the Kronecker, outer vector, and  Khatri-Rao products, respectively; $|\cdot|$ is the element-wise magnitude or absolute value; $[\![\cdot]\!]$ is the multi-linear product; $(\cdot)^{\ast}$, $(\cdot)^{T}$, $(\cdot)^{H}$, and $\|\cdot\|_{\ast}$ denote conjugate, transpose, Hermitian and nuclear norm of a matrix, respectively. Following denote operators: $\mathcal{P}_{\Gamma}\{\cdot\}$ selects only those entries of its argument that are listed within the index set $\Gamma$; $\mathcal{H}\{\cdot\}$ transforms a vector to its corresponding Hankel matrix; $\mathcal{S}_{\emptyset}\{\cdot\}$ removes the zero entries; $\text{vec}(\cdot)$ vectorizes a tensor by stacking its matrices column-wise.
\section{System Model}
\label{sec:model}
\begin{figure}
    \centering
    \includegraphics[width=6.0cm]{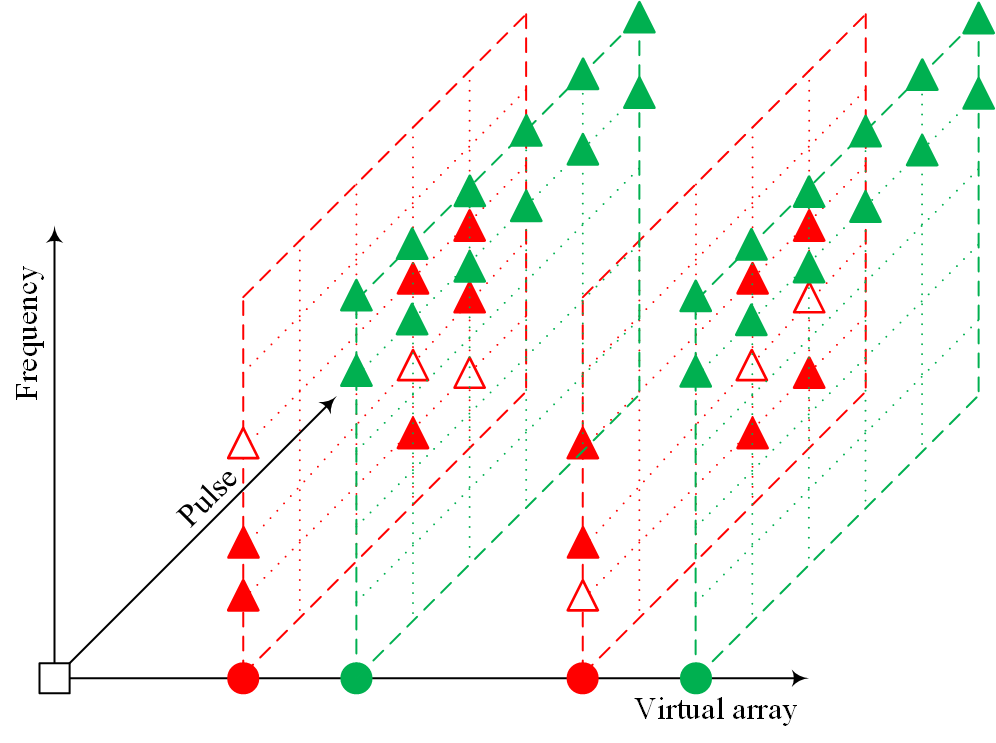}
\caption{Illustration of sub-Nyquist sampling. Different colors refer to different transmit antenna elements. The solid circles denote sampling instants along the virtual aperture. The triangles in one Frequency-Pulse plane indicate the measurements correspond to the same transmit and receive elements.}
    \label{Fig:SpatialCompression}
\end{figure}
Consider a traditional collocated MIMO radar with full Nyquist sampling. The operating wavelength of the radar is $\lambda$ and the total number of transmit and receive elements are $T$ and $R$, respectively. The MIMO antenna adopts a virtual Uniform Linear Array (ULA) structure, with receive antennas spaced $\lambda/2$ and transmit antennas spaced by $R\lambda/2$. The coherent processing of a total of $TR$ channels in the receiver creates a virtual equivalent of a phased array antenna that has $TR$ elements each spaced $\lambda/2$ distance apart.

As shown in Fig. \ref{Fig:SpatialCompression}, spatial compression is realized by a thinned array, which has $M<T$ transmit and $Q<R$ receive antennas, as in \cite{Cohen2016SUMMeR}. The $m$th transmitting antenna is located at $\xi_m\lambda/2$, and the $q$th receive antenna is located at $\zeta_q\lambda/2$, where $\xi_m$ and $\zeta_q$ are integers. The pulse train sent by the $m$th transmitting antenna over a coherent processing interval (CPI) spanning duration of $P$ pulses is
\par\noindent \small\begin{equation}
s_m(t)=\sum_{p=0}^{P-1}\delta_{m}[p]\cdot h_m(t-p\tau)e^{j2\pi f_c t}, \quad0\le t\le P\tau,
\end{equation}\normalsize
\noindent where $\tau$ denotes the pulse repetition interval (PRI), $P\tau$ is the coherent processing interval (CPI), ${{f}_{c}} = c/\lambda$ is the common carrier frequency
, $c$ is the speed of light, $\delta_m[p]=1$ or $0$ indicates whether in the $p$th PRI, the $m$th transmitter emits a pulse or not, and $\{{{h}_{m}}\left( t \right)\}_{m=0}^{M-1}$ is a set of narrowband, orthogonal pulses each with the continuous-time Fourier transform (CTFT) $H_m(\omega)=\int\limits_{-\infty}^{\infty}h_m(t)e^{-j\omega t}dt$. 

In TenDSuR, only $P_m < P$ pulses are emitted by each transmitter in a CPI and $\delta_m[p]$ can admit different values for the same $p$ and different $m$. In SUMMeR, $\delta_m[p]$ is always unity for all $m$ and $p$. We define the total duration for which the antenna elements are dedicated to a specific RF service as the aperture occupancy, $\mathrm{AO}=\sum_{m=0}^{M-1}P_m$.

Suppose there exist $L$ non-fluctuating point targets (Swerling-0 model), where the $l^{\text{th}}$ target is characterized by its complex reflectivity $\alpha_l$, range $r_l$, radial velocity $v_l$, DoA $\vartheta_l = \sin\theta_l$ ,and $\theta_l$ is the azimuth. As derived in \cite{Cohen2016SUMMeR}, the received echo of the $p$th pulse at the $q$th receiving antenna can be represented by its Fourier series, as
\par\noindent \small\begin{equation}
x_q^p(t)=\sum_{n\in\mathbb{Z}}c_q^p[n]e^{-j2\pi nt/\tau},
\end{equation}\normalsize
where for $-NT/2\leq n\leq NT/2-1$, with $N=B_h\tau$, 
\par\noindent\small
\begin{eqnarray}
\label{Eq:EchoCoefficientsPartial_scalar}
c_q^p[n]=\frac{1}{\tau}\sum_{m=0}^{M-1}\sum_{l=1}^{L}\delta_m[p]\cdot\alpha_le^{j2\pi\beta_{mq}\vartheta_l}e^{-j\frac{2\pi}{\tau}n\tau_l}e^{-j2\pi f_l^Dp\tau}H_m\left(\frac{2\pi}{\tau}n\right),
\end{eqnarray}
\normalsize
where time delay $\tau_l = 2r_l/c$ is proportional to $l$th target's range $r_l$, $f_l^D=2\frac{v_l}{c}f_c$ is the Doppler frequency, parameter ${\beta _{mq}} = ({\zeta _q} + {\xi _m})({f_m}\frac{\lambda }{c} + 1)$ is governed by the array structure.

We now apply Xampling in time, pulses and space to obtain low-rate samples of the received signal. The sampling technique is the same as in temporal sub-Nyquist radar \cite{barilan2014focusing}, except that now the samples are obtained in range, Doppler and azimuth domains. The received signal $x^p_q(t)$ is downconverted to baseband, separated into $M$ channels, aligned and normalized. The Fourier coefficients of the received signal corresponding to the channel that processes the $p$th pulse of $m$th transmitter echo at the $q$th receiver are\par\noindent\small
\begin{eqnarray}
\label{Eq:BasebandEchoCoefficients}
y_{mq}^p[k]&=&\tau H^{\ast}_m(2\pi k/\tau + f_m)c_{q}^{p}[k+f_m \tau]/|H_0(2\pi k/\tau){|^2}\nonumber\\
&=&\sum_{l=1}^{L}\delta_m[p]\cdot\alpha_le^{j2\pi\beta_{mq}\vartheta_l}e^{-j\frac{2\pi}{\tau}k\tau_l}e^{-j2\pi f_m\tau_l}e^{-j2\pi f_l^{D}p\tau}.
\end{eqnarray}\normalsize
where $-\frac{N}{2} \leq k \leq -\frac{N}{2}-1$, $f_m$ is the (baseband) carrier frequency of the $m$th transmitter and $N$ is the number of Fourier coefficients per channel. Xampling obtains a set $\mathcal{K}$ of arbitrarily chosen Fourier coefficients from low rate samples of the received channel signal such that $|\mathcal{K}| = K < N$.

Let $\mathbf{Z}^m$ be the $KQ\times P_m$ matrix with $q$th column given by the vertical concatenation of $y_{mq}^p[k]$, $k \in \mathcal{K}$, $0 \leq q \leq Q-1$ and $p \in \Pi_m$, where $\Pi_m$ is the set of pulses chosen arbitrarily for transmission by the $m$th transmit antenna out of a total of $P$ pulses such that $|\Pi_m| = P_m$. 
We can write $\mathbf{Z}^m$ as
\par\noindent \small\begin{equation}
\label{Eq:SUMMeRMode2}
\mathbf{Z}^m=(\mathbf{B}^m\otimes\mathbf{A}^m)\mathbf{X}_{\rm{D}}(\mathbf{F}^m)^T,
\end{equation}\normalsize
where $\mathbf{A}^m$ denotes the $K \times TN$ matrix whose $(k,n)$th element is $e^{- j \frac{2 \pi}{TN} \mathcal{K}_kn} e^{-j2\pi \frac{f_m}{B_h} \frac{n}{T}}$ with $\mathcal{K}_k$ the $k$th element in $\mathcal{K}$, $\mathbf{B}^m$ is the $Q \times TR$ matrix with $(q,p)$th element $e^{-j2 \pi \beta_{mq} (-1 +\frac{2}{TR}p)}$ and $\mathbf{F}^m$ denotes the $P_m \times P$ partial Fourier matrix. The matrix $\mathbf{X}_D$ is a $T^2NR \times P$ sparse matrix that contains the values $\alpha_l$ at the $L$ indices. In TenDSuR, the row size of partial Fourier matrix $\mathbf{F}^m$ can be different across all $m$. However, in SUMMeR, $\mathbf{F}^m$ is always a full Fourier matrix for all transmitters.

Vectorizing both sides of (\ref{Eq:SUMMeRMode2}) gives $\mathrm{vec}(\mathbf{Z}^m)=\mathrm{vec}((\mathbf{B}^m\otimes\mathbf{A}^m)\mathbf{X}_{\rm{D}}(\mathbf{F}^m)^T) =(\mathbf{F}^m\otimes\mathbf{B}^m\otimes\mathbf{A}^m)\mathrm{vec}(\mathbf{X}_{\rm{D}})$.
Consider a $K\times Q\times P_m$ tensor $\mathcal{Z}^m$ such that $\mathrm{vec}(\mathbf{Z}^m)=\mathrm{vec}(\mathcal{Z}^m)$. Then,
\par\noindent \small\begin{equation}
\label{Eq:TensorModel_m}
\mathcal{Z}^m=[\![\mathcal{X}; \mathbf{A}^m, \mathbf{B}^m, \mathbf{F}^m]\!],
\end{equation}\normalsize
where $\mathcal{X}\in\mathbb{C}^{TN\times TR\times P}$ such that $\mathrm{vec}(\mathbf{X}_{\rm{D}})=\mathrm{vec}(\mathcal{X})$.

The matrices $\mathbf{A}^m$, $\mathbf{B}^m$ and $\mathbf{F}^m$ are obtained by selecting $K$, $Q$ and $P_m$ rows, respectively, of their full counterparts: $TN\times TN$ matrix $\mathbf{A}$, $TR\times TR$ matrix $\mathbf{B}$ and $P \times P$ matrix $\mathbf{F}$. Hence, the partial tensor $\mathcal{Z}^m$ is obtained by selecting a total of $KQ\sum_{m=0}^{M-1}P_m$ entries from the corresponding full tensor $\mathcal{Z}=[\![\mathcal{X}; \mathbf{A}, \mathbf{B}, \mathbf{F}]\!]$ of size $TN\times TR\times P$.

For on-grid targets, the Fourier coefficients of the received signal are obtained from $\mathcal{Z}$ as
\par\noindent \small\begin{equation}
  \label{Eq:SignalModelTensor}
 \bar{\mathcal{Z}} = \mathcal{P}_{\Gamma}\{\mathcal{Z}\} = \mathcal{P}_{\Gamma}\big\{[\![\mathcal{X}; \mathbf{A}, \mathbf{B}, \mathbf{F}]\!]\big\},
\end{equation}\normalsize
where $\bar{\mathcal{Z}}$ represents the entries of $\mathcal{Z}$ specified by the set $\Gamma$ which is constructed as per the Algorithm~\ref{Alg:Construction}.
Here, the sets $\Omega$, $\Lambda$ and $\Pi$ index the rows of $\mathbf{A}^m$, $\mathbf{B}^m$ and $\mathbf{F}^m$ selected from $\mathbf{A}$, $\mathbf{B}$ and $\mathbf{F}$, respectively.
%
\begin{algorithm}
  \caption{The construction of index sets $\Gamma$, $\Omega$, $\Lambda$, and $\Pi$.}
  \label{Alg:Construction}
  \begin{algorithmic}[1]
    \Require
     $H_m(2\pi (k)/\tau)$;
    \Ensure
      $\Gamma$, $\Omega$, $\Lambda$, $\Pi$, ;
    \State  $\Gamma={\emptyset}$, $\Omega={\emptyset}$, $\Lambda={\emptyset}$, $\Pi={\emptyset}$, ;

\For{$m=0$ to $M-1$} 
\For{$q=0$ to $Q-1$}  
\For{$p=0$ to $P-1$}
\For{$k\in\mathcal{K}$}
\If {$\delta_m[p]\cdot|H_m(2\pi k/\tau)| \neq 0$ }
 \State  $\Gamma=\Gamma \cup \{k+f_m \tau,\xi_m+\zeta_q,p\}$.
$\Omega=\Omega \cup \{k+f_m \tau\}$, $\Lambda = \Lambda \cup \{\xi_m+\zeta_q\}$, $\Pi=\Pi \cup \{p\}$,
\EndIf
\EndFor
\EndFor
\EndFor
\EndFor
   \end{algorithmic}
\end{algorithm}

For continuous-valued parameters, the entries of $\bar{\mathcal{Z}}$ are 
\par\noindent \small\begin{equation}
\label{Eq:EchoCoefficientsPartial}
[\bar{\mathcal{Z}}]_{k+f_m\tau,\xi_m+\zeta_q,p}=\sum_{l=1}^{L}\alpha_l \cdot e^{j2\pi\beta_{mq} \vartheta_l} e^{-j\frac{2\pi}{\tau}k\tau_l } e^{-j2\pi f_m\tau_l}e^{-j2\pi f_l^Dp\tau}.
\end{equation}\normalsize
\section{Target Recovery}
\label{sec:algorithm}
We first derive the number of samples needed for perfect recovery of $\{\alpha_l, \vartheta_{l}, \tau_{l}, f^D_{l}\}_{l=1}^L$ or $\mathcal{X}$. In the results below, the total number of transceiver frequencies, antennas, and pulses equals the number of $\bar{\mathcal{Z}}$'s non-zero matrix slices obtained by fixing the first, second, and third index, respectively.
\begin{theorem}
\label{TH: MiniContin}
The minimal total number of antenna channels, transceiver frequencies, and pulses required for perfect recovery of $L$ off-grid targets in a noiseless setting are each no less than $2L$.
\end{theorem}
\begin{IEEEproof}  We first prove the necessary condition for the overall number of pulses. The cases for overall number of antenna channels and transceiver frequency points are similar. Consider the case where all targets have identical DoAs and ranges, as $\vartheta_l=\vartheta_0$, $\tau_l=\tau_0$. Then, from (\ref{Eq:BasebandEchoCoefficients}),
\begin{eqnarray}
\label{Eq:EchoCoefficientsIdential}
&&e^{j2\pi\tau_{0} (f_m\tau+k)/\tau}e^{j2\pi\vartheta_{0}(\xi_m+\zeta_q)}\tau H^{\ast}_m(2\pi k/\tau+f_m)c_{q}^{p}[k+f_m \tau] = \nonumber\\
&&|H_0(2\pi k /\tau)|^2\sum_{l=1}^{L}\delta_m[p]\cdot\alpha_l  e^{-j2\pi f_l^D p}.
\end{eqnarray}
Summing over $m=0, \dots, M-1$, $q=0, \dots, Q-1$, and $k\in \mathcal{K}$ on both sides yields
\par\noindent \small\begin{equation}
\label{Eq:EchoCoefficientsEqual}
z'[p]=\sum_{l=1}^{L}\alpha_l  e^{-j2\pi f_l^D p},
\end{equation}\normalsize
where
\begin{flalign}
\label{Eq:EchoCoefficientsEqualDef}
z'[p]=&\frac{\sum_{m=0}^{M-1}\sum_{q=0}^{Q-1}\sum_{k\in \mathcal{K}}e^{j2\pi\tau_{0} (f_m\tau+k)/\tau}e^{j2\pi\vartheta_{0}(\xi_m+\zeta_q)}H^{\ast}_m(2\pi k/\tau+f_m)}{Q\sum_{m=0}^{M-1}\sum_{k\in \mathcal{K}}\delta_m[p]\cdot|H_0(2\pi k/\tau)|^2 }\nonumber\\
&\cdot c_{q}^{p}[k+f_m \tau].
\end{flalign}
In (\ref{Eq:EchoCoefficientsEqualDef}), if there exists a pulse index $p'$ such that for all $k\in \mathcal{K}$, $m=0, \dots, M-1 $, the term $\delta_m[p']\cdot|H_0(2\pi k/\tau)|^2$ in the denominator vanishes, then $z'[p']$ is undefined. Then, according to FRI theory, there are only $2L$ degrees of freedom in (\ref{Eq:EchoCoefficientsEqual}) thereby requiring no less than $2L$ overall pulses for a successful recovery of $L$ targets.
\end{IEEEproof}

\begin{theorem}
\label{TH: MiniDiscre}
The minimal total number of antenna channels, transceiver frequencies, and pulses required for perfect recovery of $\mathcal{X}$ with $L$ on-grid targets in a noiseless setting are each no less than $2L$.
\end{theorem}
\begin{IEEEproof} 
For all $k\in \mathcal{K}, m=0, \dots, M-1, p=0,\dots P-1$, if $|H_0(2\pi k/\tau)|^2\neq 0$, then
he signal model in (\ref{Eq:SignalModelTensor}) can be unfolded as the following $KQ\sum_{m=0}^{M-1}P_m$ equations
\par\noindent \small\begin{equation}
[\bar{\mathcal{Z}}]_{k,\xi_m+\zeta_q,p}=\left(([\mathbf{F}^T]_p)^T\otimes ([\mathbf{B}^T]_{\xi_m+\zeta_q})^T\otimes ([\mathbf{A}^T]_{k})^T\right)\rm{vec}(\mathcal{X}).
\end{equation}\normalsize
As per the FRI theory and Lemma 1 in \cite{Cohen2016SUMMeR}, the sparse tensor $\mathcal{X}$ can be perfect recovered provided
\par\noindent \small\begin{equation}
\min\left\{spark(([\mathbf{F}^T]_{\Pi})^T), spark(([\mathbf{A}^T]_{\Omega})^T),spark(([\mathbf{B}^T]_{\Lambda})^T)\right\}> 2L.\nonumber
\end{equation}\normalsize 
As per $\Omega$, $\Lambda$, and $\Pi$, if the overall number of tranceived frequency points is larger than $2L$, then $ spark(([\mathbf{A}^T]_{\Omega})^T)>2L$; if the overall number of antenna channels is larger than $2L$, then $spark(([\mathbf{B}^T]_{\Lambda})^T)>2L$; and if the overall number of pulses is larger than $2L$, then $spark(([\mathbf{F}^T]_{\Pi})^T)>2L$.
\end{IEEEproof} 

The crucial difference between the above two theorems and their counterparts Theorems 3 and 4 in \cite{Cohen2016SUMMeR}, is that in SUMMeR, all transmitters have to transmit no less than $2L$ pulses in one CPI. However, in TenDSuR, we need at least $2L$ pulses in a CPI but only one transmitter need to be active. This significantly reduces the AO of an RF service. As the reserved antenna elements can be used for other services, the efficiency of the multi-function RF aperture is remarkably improved.

For continuous target parameters, one could estimate the unknown parameters via TC \cite{gandy2011tensor} and parallel factor analysis (PARAFAC)\cite{Sidiropoulos2017Tensor} or higher order harmonic retrieval algorithms\cite{Haardt2008Higher}. In this work, we utilize the Hankel Matrix nuclear norm
Regularized low-CP-rank Tensor Completion (HMRTC) introduced in \cite{Ying2017Hankel}, which explores both the low-CP-rank and the Vandermonde structure of the tensor ${\mathcal{Z}}$, to recovery the targets' DoAs, ranges, and velocities:
\par\noindent \small\begin{align}
\label{Eq:HMRTC}
\{\hat{\mathbf{a}}_{l}, \hat{\mathbf{b}}_{l}, \hat{\mathbf{f}}_{l}\}_{l=1}^{L}=&\arg\min_{\{\mathbf{a}_{l}, \mathbf{b}_{l}, \mathbf{f}_{l}\}_{l=1}^{L}} \sum_{l=1}^{L}
\|\mathcal{H}\{\mathbf{a}_{l}\}\|_{\ast}+\|\mathcal{H}\{\mathbf{b}_{l}\}\|_{\ast}+\|\mathcal{H}\{\mathbf{f}_{l}\}\|_{\ast}\nonumber\\
&+\frac{\mu}{2}\left\|\mathcal{P}_{\Gamma}\left\{\sum_{l=1}^{L} {{\bf{a}}_{l}} \circ {{\bf{b}}_{l}} \circ {{\bf{f}}_{l}}\right\}-\mathcal{P}_{\Gamma}\{\mathcal{Z}\}\right\|_2^2,
\end{align}\normalsize
where $\mu$ is a regularization parameter \cite{Ying2017Hankel} and, as per 
(\ref{Eq:EchoCoefficientsPartial}),
\par\noindent\small\begin{equation}
\begin{array}{l}
{{\bf{a}}_l} = {\left[ {1,{e^{ - j2\pi \frac{{{\tau _l}}}{\tau }}}, \ldots ,{e^{ - j2\pi (TN - 1)\frac{{{\tau _l}}}{\tau }}}} \right]^T},\\
{{\bf{b}}_l} = {\left[ {1,{e^{ - j2\pi ({f_m}\frac{\lambda }{c} + 1){\vartheta _l}}}, \ldots ,{e^{ - j2\pi (TR - 1)({f_m}\frac{\lambda }{c} + 1){\vartheta _l}}}} \right]^T},\\
{{\bf{f}}_l} = {\left[ {1,{e^{ - j2\pi f_l^D\tau }}, \ldots ,{e^{ - j2\pi (P - 1)f_l^D\tau }}} \right]^T}.
\end{array}
\end{equation}\normalsize
The target parameters are obtained after recovering $\{\hat{\mathbf{a}}_{l}, \hat{\mathbf{b}}_{l}, \hat{\mathbf{f}}_{l}\}_{l=1}^{L}$. For on-grid target parameters, the $\mathcal{X}$ or $\mathbf{x}$ is retrieved by solving the sparse recovery problem:
\par\noindent \small\begin{equation}
\label{Eq:DefTensorL0MinNoisy}
\min_{\mathcal{X}}\|\mathcal{X}\|_{0}\phantom{1}\text{subject to}\phantom{1} \|\mathcal{P}_{\Gamma}\{[\![\mathcal{X}; \mathbf{A}, \mathbf{B}, \mathbf{F}]\!]\}-\mathcal{P}_{\Gamma}\{\mathcal{Z}\}\|_2\leq \varepsilon,\end{equation}\normalsize
where $\varepsilon$ is the error threshold determined by the noise power. Among several CS algorithms to solve this problem \cite{eldar2012compressed}, we use Orthogonal Matching Pursuit (OMP) \cite{eldar2015sampling}, popular for its good trade-off between recovery accuracy and computational load. Our Algorithm~\ref{Alg:OMP} is the tensor version of OMP. Here,
\par\noindent \small\begin{equation}
\label{Eq:SupportObservingMatrix}
\mathbf{U}_{\Phi_i}\triangleq[\bar{\mathbf{F}}]_{\Phi_{i}(3)}\diamond[\bar{\mathbf{B}}]_{\Phi_{i}(2)}\diamond[\bar{\mathbf{A}}]_{\Phi_{i}(1)},
\end{equation}\normalsize
where $\bar{\mathbf{A}}=[\mathbf{A}_0^T, \mathbf{A}_1^T,\dots, \mathbf{A}_{M-1}^T]^T$, $\bar{\mathbf{B}}=[\mathbf{B}_0^T, \mathbf{B}_1^T,\dots, \mathbf{B}_{M-1}^T]^T$, $\bar{\mathbf{F}}=[\mathbf{F}_0^T, \mathbf{F}_1^T,\dots, \mathbf{F}_{M-1}^T]^T$, and $\Phi_i$ is a set whose entries are triples of integers, and $\Phi_{i}(j)$ $(j=1, 2, 3)$ is a set composed of all the $j$th integer in the triples. $\Omega_{\Gamma}$ is a set of integers which has the same cardinality as $\Gamma$, and the $k$th entry of  $\Omega_{\Gamma}$ is $[\Gamma(1)]_k\cdot (TR)P + [\Gamma(2)]_k\cdot P + [\Gamma(3)]_k$, where the definition of $\Gamma(j)$ is similar as that of $\Phi_{i}(j)$. Once the tensor $\mathcal{X}$ and its support set $\Phi$ are recovered, the target parameters are estimated as $
\phantom{1}\hat{\tau}_l=\frac{\tau}{TN}\phi_{l}(1)$, $\phantom{1}\hat{\vartheta}_l=\frac{2}{TR}\phi_{l}(2)$ and $\phantom{1}\hat{f}^D_l=\frac{1}{P\tau}\phi_{l}(3).$
\begin{algorithm}
  \caption{Target Recovery in TenDSuR through OMP}
  \label{Alg:OMP}
  \begin{algorithmic}[1]
    \Require
     $\Gamma$,  $\bar{\mathcal{Z}}$, $L$;
    \Ensure
      Estimated support $\hat{\Phi}$ of $\mathcal{X}$, sparse tensor estimate $\hat{\mathcal{X}}$;
    \State (\textit{Initialization}): $\mathcal{R}=\bar{\mathcal{Z}}$, $\Phi_0={\emptyset}$, $i=1$;
    \State (\textit{Projection}): $\mathcal{Y} = [\![\mathcal{R}; \mathbf{A}^H, \mathbf{B}^H, \mathbf{F}^H]\!]$;
   \State (\textit{Support set augmentation}): $\Phi_i=\Phi_{i-1}\cup \{\phi_i\}$, where
   \par\noindent \small\begin{equation}
   \phi_i=\big\{\phi_{i}(1),\phi_{i}(2),\phi_{i}(3)\big\} =\arg\max_{\{s_1,s_2,s_3\}}\big|[\mathcal{Y}]_{s_1,s_2,s_3}\big|;\nonumber
   \end{equation}\normalsize
   \State (\textit{Signal estimation}): $
  \bm{\alpha}_i=\big(\mathbf{U}_{\Phi_i}^H\mathbf{U}_{\Phi_i}\big)^{-1}\mathbf{U}_{\Phi_i}^H\mathcal{S}_{\emptyset}\left\{\mbox{vec}\big(\bar{\mathcal{Z}}\big)\right\};\nonumber
   $
   \State (\textit{Compute residual}):
   
   $
   \mathcal{R} = \bar{\mathcal{Z}}-\mathcal{P}_{\Gamma}\left\{\sum_{l=1}^{i}[\bm{\alpha}_i]_l \cdot \left([\mathbf{A}]_{\phi_{l}(1)}\circ[\mathbf{B}]_{\phi_{l}(2)}\circ[\mathbf{F}]_{\phi_{l}(3)}\right)\right\}$;
   \State (\textit{Iteration}): if $i < L$, $i=i+1$ and go to Step 2, else stop;
   \State (\textit{Output}): $\hat{\Phi} = \Phi_L$; $\hat{\mathcal{X}}\in \mathbb{C}^{TN\times TR\times P}$, where
   \par\noindent \small\begin{equation}
   [\hat{\mathcal{X}}]_{\phi_{l}(1),\phi_{l}(2),\phi_{l}(3)} =\left\{\begin{array}{cc}
   [\bm{\alpha}]_l, &l=1,2,\dots, L,\nonumber\\
   0, &\mbox{otherwise.}
   \end{array}   \right.
   \end{equation}\normalsize
   \end{algorithmic}
\end{algorithm}
\section{Numerical Experiments}
\label{sec:results}
In all our numerical experiments, radar with full antenna aperture $TR=20$, $\lambda = 0.03$ m, $P=16$, and $\tau = 0.016$ ms was compressed by a thinned MIMO array with $M = 2$ transmit and $Q = 5$ receive elements with locations $\{0,\phantom{1}\lambda/2\}$ and $\{\lambda,\phantom{1}5\lambda/2,\phantom{1}11\lambda/2,\phantom{1}13\lambda/2,\phantom{1}15\lambda/2\}$, respectively. 
The native range resolution corresponding to the full bandwidth is $150$ m. Thus, $TN=16$. 
We assumed additive white Gaussian noise was present at each receiver with identical noise power. We consider the matched filter definition of the signal-to-noise-ratio, $
 {\rm{SNR}} = \frac{\left(\sum_{m=0}^{M-1}\sum_{k\in\mathcal{K}}|H_0(2\pi k/\tau)|\right)^2}{\sigma^2\cdot KQ\sum_{m=0}^{M-1}P_m}$,
where $\sigma ^2$ denotes the noise power corresponding to the bandwidth $1/\tau$ of a single receiver.
\begin{figure}
    \centering
    \includegraphics[width=8.5cm]{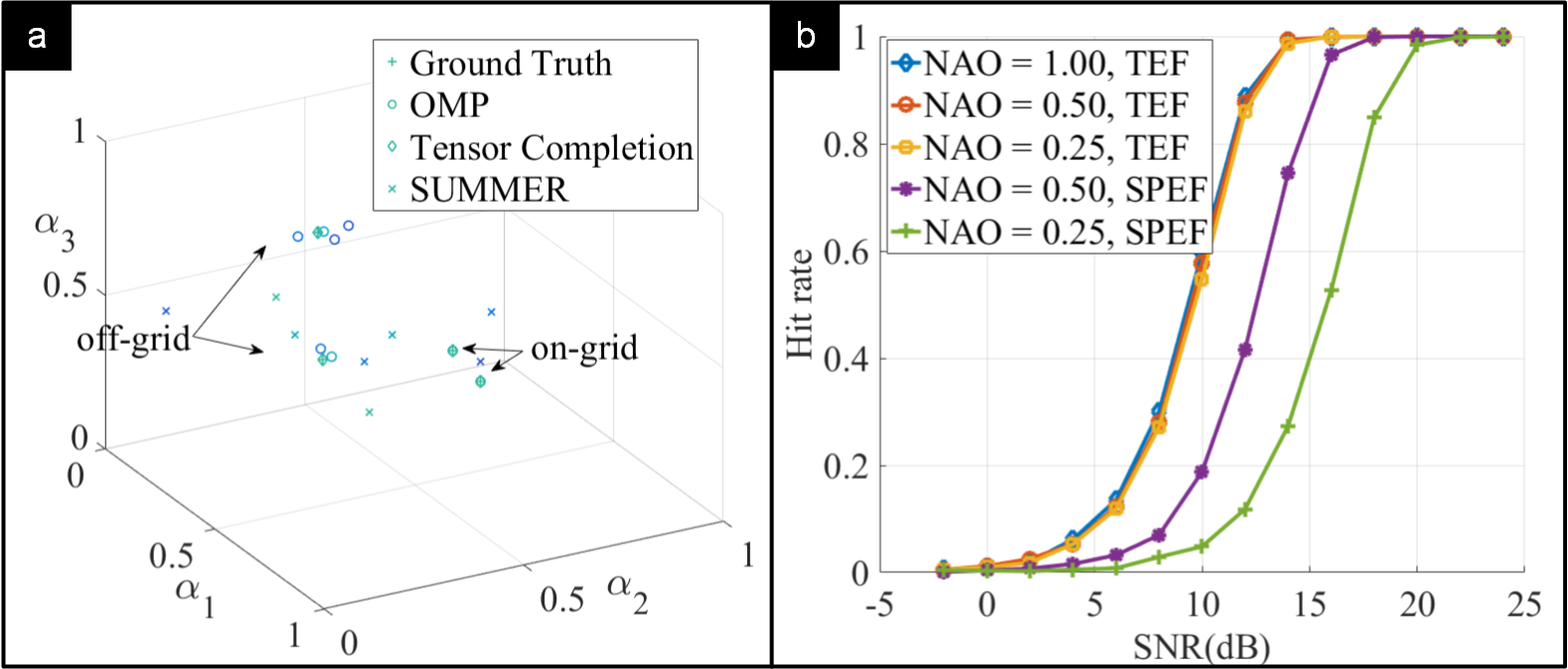}
    \caption{(a) The recovered normalized delay-Doppler-DoA map. Detections via tensor-OMP, TC and SUMMeR are indicated by circles, diamonds, and oblique crosses respectively. The crosses indicate true location of the targets. (b) Hit rate comparison under different SNRs and compression ratios. (TEF for total energy fixed and SPEF for single pulse energy fixed.)}
    \label{Fig:Fourtarget_DifferentSNR}
\end{figure}

First, we provide an example of target detection with S-S-D-T compression. We consider a target scene with $L = 4$ targets distributed in the direction-range-velocity observation scene, whose reflectivities were all set to unity (see Fig.~\ref{Fig:Fourtarget_DifferentSNR}(a)). Two of the targets were placed on the grid while the others were off-grid. In every PRI, the 
number of frequency points transmitted by each transmitter were $4$, i.e. a quarter of the whole bandwidth.
The frequency points of the two transmitters were kept different from each other to ensure waveform orthogonality. To quantify the effect of dilution in the Doppler domain, we define the Normalized Aperture Occupancy (NAO) as the AO divided by the whole transmit pulse product: $
\mathrm{NAO}=\frac{\sum_{m=0}^{M-1}P_m}{M P}$. In this experiment, each transmitter sent only $4$ pulses implying $\mathrm{NAO}=0.25$. In comparison, the NAO of SUMMeR must be no less than $0.5$. 
We apply the hit-or-miss criterion as the performance metric to compare tensor-OMP and TC. Here, a ``hit'' occurs when the estimated target is within one Nyquist bin (defined as $2/(TR)$, $\tau/(TN)$, and $1/(P\tau)$ for the DoA, delay and Doppler, respectively - from the true target location. In Fig.~\ref{Fig:Fourtarget_DifferentSNR}(a), the on-grid targets are successfully detected by both OMP and TC. However, only TC showed successful detections of off-grid targets at the exact location while OMP found several solutions many bins away from the true target location. The Doppler-focusing-based SUMMeR recovery algorithm \cite{Cohen2016SUMMeR} fails to recover most of the targets.

Next, we placed $L = 2$ targets on the scene. Each pulse contained $4$ frequency points. The hit rate was computed over 10,000 Monte-Carlo trials 
(see Fig. \ref{Fig:Fourtarget_DifferentSNR}(b)). We note that when the total transmitted energy is the same, the reduction of AO does not significantly influence target detection. On the other hand, when the transmit energy in each pulse is fixed, performance deteriorates and the hit rate curve moves to approximately $3$ dB and $6$ dB to the right for $\mathrm{NAO}=0.5$ and $0.25$, respectively.


\clearpage
\balance
\bibliographystyle{IEEEtran}
\bibliography{refs}

\end{document}